\documentclass[onecolumn, amsmath, showpacs,11pt]{revtex4}
\usepackage{graphicx}
\usepackage{dcolumn}
\usepackage{bm}
\usepackage{color}
\usepackage{slashed}
\begin{document}
\newcommand{\hs}{\hspace*{0.5cm}}
\newcommand{\vs}{\vspace*{0.5cm}}
\newcommand{\be}{\begin{equation}}
\newcommand{\ee}{\end{equation}}
\newcommand{\bea}{\begin{eqnarray}}
\newcommand{\eea}{\end{eqnarray}}
\newcommand{\ben}{\begin{enumerate}}
\newcommand{\een}{\end{enumerate}}
\newcommand{\bde}{\begin{widetext}}
\newcommand{\ede}{\end{widetext}}
\newcommand{\nn}{\nonumber}
\newcommand{\crn}{\nonumber \\}
\newcommand{\Tr}{\mathrm{Tr}}
\newcommand{\non}{\nonumber}
\newcommand{\noi}{\noindent}
\newcommand{\al}{\alpha}
\newcommand{\la}{\lambda}
\newcommand{\bet}{\beta}
\newcommand{\ga}{\gamma}
\newcommand{\va}{\varphi}
\newcommand{\om}{\omega}
\newcommand{\pa}{\partial}
\newcommand{\+}{\dagger}
\newcommand{\fr}{\frac}
\newcommand{\bc}{\begin{center}}
\newcommand{\ec}{\end{center}}
\newcommand{\Ga}{\Gamma}
\newcommand{\de}{\delta}
\newcommand{\De}{\Delta}
\newcommand{\ep}{\epsilon}
\newcommand{\varep}{\varepsilon}
\newcommand{\ka}{\kappa}
\newcommand{\La}{\Lambda}
\newcommand{\si}{\sigma}
\newcommand{\Si}{\Sigma}
\newcommand{\ta}{\tau}
\newcommand{\up}{\upsilon}
\newcommand{\Up}{\Upsilon}
\newcommand{\ze}{\zeta}
\newcommand{\ps}{\psi}
\newcommand{\Ps}{\Psi}
\newcommand{\ph}{\phi}
\newcommand{\vph}{\varphi}
\newcommand{\Ph}{\Phi}
\newcommand{\Om}{\Omega}

\title{Unifying the electroweak and $B-L$ interactions}       

\author{P. V. Dong}
\email {pvdong@iop.vast.ac.vn} \affiliation{Institute of Physics, Vietnam Academy of Science and Technology, 10 Dao Tan, Ba Dinh, Hanoi, Vietnam}
\date{\today}

\begin{abstract}
We argue that the gauge symmetry which includes $SU(3)_L$ as a higher weak-isospin symmetry is manifestly given by $SU(3)_C\otimes SU(3)_L\otimes U(1)_X\otimes U(1)_N$, where the last two factors determine the electric charge and $B-L$, respectively. This theory not only provides a consistent unification of the electroweak and $B-L$ interactions, but also gives insights in dark matter, neutrino masses, and inflation. The dark matter belongs to a class of new particles that have wrong $B-L$ numbers, and is stabilized due to a newly-realized $W$-parity as residual gauge symmetry. The $B-L$ breaking field is important to define the $W$-parity, seesaw scales, and inflaton. Furthermore, the number of fermion generations and the electric charge quantization are explained naturally. We also show that the previous 3-3-1 models are only an effective theory as the $B-L$ charge and the unitarity argument are violated. This work substantially generalizes our recently-proposed 3-3-1-1 model.                  
\end{abstract}


\maketitle

\section{Introduction} 

The standard model is incomplete since it leaves many striking features of the physics of our world unanswered \cite{pdg}. The leading questions perhaps include the neutrino masses, dark matter, matter-antimatter asymmetry, and cosmic inflation. The standard model also cannot explain why there are only three fermion generations and what makes the electric charges be quantized.              

The most popular solutions for the observed, small neutrino masses could be the seesaw mechanisms \cite{seesaw}. Interestingly, they also lead to leptogenesis processes that address the matter-antimatter asymmetry. The crucial keys of the type I and type II seesaw mechanisms are at the seesaw scales, which keep the small neutrino masses. However, at present they have been less understood. What is their origin? Which is the physics behind? Can the seesaw scales be related? Further, the generation of the thermal dark matter relic density implies the existence of a weakly-interacting massive particle (WIMP) \cite{dmreview}. Many simple extensions of the standard model provide WIMPs. But, what is the WIMP nature? Why is it stabilized? Can the WIMP that is bounded below some hundreds of TeV be correlated to the inflationary dynamics at the grand unification scale. Could the seesaw and inflationary scales be common?       

As an attempt to address those questions, this work is a substantial generalization of a recently-proposed $SU(3)_C\otimes SU(3)_L\otimes U(1)_X\otimes U(1)_N$ (3-3-1-1) gauge model \cite{d3311}. We will strictly derive the 3-3-1-1 gauge symmetry, along with the introductory of the most general fermion content. For this aim, we start from $SU(3)_L$, a higher weak-isospin symmetry directly extended from $SU(2)_L$, which is best known for solving the number of observed fermion generations. To preserve the electric charge, baryon and lepton number symmetries, the complete gauge symmetry must be $SU(3)_L\otimes U(1)_X\otimes U(1)_N$ (besides the $SU(3)_C$ color group), where $X$ and $N$ define the electric charge and baryon minus lepton number, respectively. The general fermion content is free from all the anomalies, where the most new fermions (including the non-Hermitian gauge bosons) have new, characteristic $B-L$ quantum numbers.   

The scalar sector is introduced and the 3-3-1-1 symmetry breaking is discussed. The new model yields a discrete symmetry, called $W$-parity (although it is actually larger than $Z_2$), as a remnant of the gauge symmetry, which separates the model particles into two classes, normal particles and wrong particles. The wrong particles transform nontrivially under the $W$-parity, and are only coupled in pairs in interactions, similarly to the superparticles in supersymmetry. The $W$-parity makes some wrong lepton or baryon particle stable, providing dark mater candidates. The neutrino masses are generated as a result of the gauge symmetry breaking, where the seesaw mechanisms are naturally realized. The model also provides inflaton as the dynamics of $B-L$ breaking as well as leptogenesis processes automatically. The hints of the electric charge quantization are shown. The gauge bosons are identified, and the corresponding constraints are given. The unitarity of the model as well as the previous theories is also investigated.    

The rest of this work is organized as follows. In Section \ref{model} we construct the model. Here, the dark matter, neutrino masses, and the quantization of charges are also discussed. Section \ref{gauge} is devoted to the gauge bosons and some constraints. The unitarity is considered in Section \ref{unitaritys}. The cosmological inflation and letogenesis are discussed in Section \ref{infbasym}. We summarize our results and conclude this work in Section \ref{conclusion}.                          

\section{\label{model} Proposal of the model}

\subsection{3-3-1-1 symmetry and fermion content}   

The first observation is that in the standard model the $[SU(2)_L]^3$ anomaly always vanishes, $\Tr[\{T_a,T_b\}T_c] = 0$, for any chiral fermion representation, where $T_a\ (a=1,2,3)$ indicate $SU(2)_L$ generators. Let $SU(2)_L$ be enlarged to $SU(3)_L$, a higher weak-isospin symmetry. As a result, the corresponding anomaly $[SU(3)_L]^3$ does not vanish, $A_{ijk} \equiv \Tr[\{T_i,T_j\}T_k]\neq 0$, for complex representations, where $T_i\ (i=1,2,3,...,8)$ denote $SU(3)_L$ generators. This subsequently give constraints on the new fermion content \cite{anomaly}. The new gauge symmetry must span $SU(3)_C\otimes SU(3)_L$, where the first factor is ordinary color group.   

The fundamental representations of $SU(3)_L$ are decomposed as $3=2 \oplus 1$ and $3^*=2^*\oplus 1$ under $SU(2)_L$. Therefore, all the (left-handed) fermion doublets of $SU(2)_L$ will be embedded into $3$ or $3^*$, where for the latter the antidoublets take the form $(f_2\ -f_1)$, provided that $(f_1\ f_2)$ is a doublet. We also suppose that all the (right-handed) fermion singlets of $SU(2)_L$ by themselves transform as corresponding singlets of $SU(3)_L$. Because of $A_{ijk}(3^*)=-A_{ijk}(3)$, the $[SU(3)_L]^3$ anomaly is cancelled out if the number of $3$ equals to the number of $3^*$ (where the color number must be appropriately counted). Therefore, the fermion representations under $SU(3)_L$ are arranged as  
\bea && \psi_{aL}\equiv 
\left(\begin{array}{c}
\nu_{aL}\\
e_{aL}\\
k_{aL}\end{array}\right)\sim 3, \hs Q_{3 L}\equiv 
\left(\begin{array}{c}
u_{3L}\\
d_{3L}\\
j_{3L}\end{array}\right)\sim 3,\hs Q_{\al L}\equiv 
\left(\begin{array}{c}
d_{\al L}\\
-u_{\al L}\\
j_{\al L} \end{array}\right)\sim 3^*,\\ 
&& \nu_{aR},\ e_{aR},\ k_{a R},\ u_{a R},\ d_{a R},\ j_{a R}\sim 1, \eea    
where $a=1,2,3$ and $\al =1,2$ are generation indices, and $\nu_{aR},\ k_a,\ j_a$ are new particles, which are $SU(2)_L$ singlets added to complete the representations.   

As a matter of fact, we possibly have a special case where $k_{a}$ are excluded (not needed). Instead, the third components of $\psi_a$ (like the 1's in the above decompositions) can be assigned by either $e_{aR}$ or $\nu_{aR}$, called minimal versions. Namely, $k_{aR}$ are suppressed, while $k_{aL}$ are replaced by either $(e_{aR})^c$ or $(\nu_{aR})^c$, where ``$c$'' indicates the charge conjugation, $(f_R)^c\equiv C\bar{f_R}^T=(f^c)_L$, as usual. However, this does not work for the case of quarks because $SU(3)_L$, $SU(3)_C$, and the space-time symmetry commute. Hence, the introductory of $j_a$ is necessary. Furthermore, the results obtained below generally apply for all cases. A direct consequence of the above proposal is that the number of fermion generations must equal to the fundamental-color number \cite{nfg,decq}.        

Since $k_a$ are unknown, let their electric charges be $q$. Furthermore, the electric charge operator $Q$ does not commute and noncloses algebraically with $SU(3)_L$. Indeed, for a lepton triplet, we have $Q=\mathrm{diag}(0,-1,q)$ which is generally not commuted with $T_i=\fr{1}{2}\la_i$ for $i=1,2,4,5,6,7$:
\bea \left[Q, T_1\pm i T_2\right] &=& \pm (T_1\pm i T_2),\crn
\left[Q, T_4\pm i T_5\right] &=& \mp q (T_4\pm i T_5),\crn
\left[Q, T_6\pm i T_7\right] &=& \mp (1+q) (T_6\pm i T_7). \eea
The algebraic noncloseness results from the fact that if $Q$ is some generator of $SU(3)_L$, we have a combination $Q=x_i T_i$, which is invalid for $u_R$, $d_R$, even for some triplets/antitriplets since $\mathrm{Tr}Q=0$. In other words, $Q$ and $T_i$ by themselves do not make a symmetry. 

To have a closed algebra, we introduce an Abelian charge $X$ so that $Q$ is a residual charge of closed group $SU(3)_L\otimes U(1)_X$, i.e. $Q=x_i T_i + y X$. Acting $Q$ on a lepton triplet, we obtain
\be Q=T_3+\beta T_8+X, \label{eco}\ee where $\beta = -(1+2q)/\sqrt{3}$, and the weak hypercharge is identified as $Y=\beta T_8+X$. It is easily obtained the electric charges of $j_a$, $Q(j_3)=\fr 2 3+q$ and $Q(j_\al)=-\fr 1 3 -q$. Remark: since $T_{3,8}$ are gauged charges, $Q$ and $X$ must be gauged charges. This is a consequence of the non-commutation of $Q$ and $SU(3)_L$. At this stage, we conclude that the gauge symmetry of the theory must span $SU(3)_C\otimes SU(3)_L\otimes U(1)_X$. It has been extensively studied in the literature \cite{331m,331r}. 

Since $k_a$ are unknown, let their $B-L$ charges be $n$. We also assume that $B-L$ is conserved, which is actually approved by the standard model and experiments \cite{pdg}. Similarly to $Q$, we can show that $B-L$ does not commute and non-closes algebraically with $SU(3)_L$, which differs from the standard model. Indeed, for a lepton triplet, $B-L=\mathrm{diag}(-1,-1,n)$, and we have
\bea
\left[B-L, T_4\pm i T_5\right] &=& \mp (1+n) (T_4\pm i T_5),\crn
\left[B-L, T_6\pm i T_7\right] &=& \mp (1+n) (T_6\pm i T_7), \eea which non-vanish since $n$ can in principle be arbitrary.    
Even for the minimal versions aforementioned, the non-commutation is explicitly hinted due to $n=1$, thus $1+n\neq 0$. Also, if $B-L$ is algebraically closed with $SU(3)_L$, it yields $B-L=a_i T_i$ which is incorrect for the right-handed fermions as well as for some triplets/antitriplets due to $\Tr(B-L)=0$. Therefore, an Abelian charge $N$ must be imposed so that $B-L$ is a residual charge of $SU(3)_L\otimes U(1)_N$, $B-L=a_i T_i+ b N$. Acting on a lepton triplet, it follows
\be B-L=\beta' T_8+N,\label{blo}\ee where $\beta'=-2(1+n)/\sqrt{3}$. It is easily identified the $B-L$ charges of $j_a$, $[B-L](j_3)=\fr 4 3 +n$ and $[B-L](j_\al)=-\fr 2 3 -n$. Similarly to $Q$ and $X$, the charges $B-L$ and $N$ must be gauged, because $T_8$ is gauged, which is a consequence of the $B-L$ and $SU(3)_L$ non-commutation. Note that $N$ cannot be identified as $X$ since they generally differ for the right-handed fermions and for the triplets/antitriplets. Hence, they are independent charges as the charges $B-L$ and $Q$ do. 

To summarize, the gauge symmetry of the theory is manifestly given as \be SU(3)_C\otimes SU(3)_L\otimes U(1)_X\otimes U(1)_N,\ee which is called 3-3-1-1 for short. It is noteworthy that the new weak-isospin theory, $SU(3)_L$, contains in it two conserved, non-commutative charges, $Q$ and $B-L$, and their algebraic closure yields the 3-3-1-1 gauge model, which describes the strong, electroweak and $B-L$ interactions. Interestingly enough, the last two interactions (electroweak and $B-L$) are unified in the same manner as those in the electroweak theory. In Appendix \ref{appalge}, we present another approach which comes the same conclusion of the 3-3-1-1 gauge symmetry.

The fermion multiplets possess the following quantum numbers 
\bea && \psi_{aL}\sim \left(1,3,\fr{-1+q}{3},\fr{-2+n}{3}\right),\hs Q_{3L}\sim \left(3,3,\fr{1+q}{3},\fr{2+n}{3}\right),\hs Q_{\al L}\sim \left(3,3^*,-\fr{q}{3},-\fr{n}{3}\right),\crn
&& \nu_{aR}\sim \left(1,1,0,-1\right),\hs e_{aR}\sim (1,1,-1,-1),\hs k_{aR}\sim (1,1,q,n),\hs u_{a R}\sim \left(3,1,\fr 2 3,\fr 1 3\right), \label{ddnng} \\ 
&& d_{aR}\sim \left(3,1,-\fr 1 3,\fr 1 3\right),\hs j_{3R}\sim \left(3,1,\fr 2 3 +q,\fr 4 3 +n\right),\hs j_{\al R}\sim \left(3,1,-\fr 1 3 -q,-\fr 2 3 -n\right),\nn \eea which are given upon the 3-3-1-1 gauge symmetries, respectively. The fermion content as given is free from all the anomalies. Indeed, what concerned is the following nontrivial anomalies, $[SU(3)_C]^2U(1)_X$, $[SU(3)_C]^2U(1)_N$, $[SU(3)_L]^2U(1)_X$, $[SU(3)_L]^2U(1)_N$, $[\mathrm{Gravity}]^2U(1)_X$, $[\mathrm{Gravity}]^2U(1)_N$, $[U(1)_X]^2U(1)_N$, $U(1)_X[U(1)_N]^2$, $[U(1)_X]^3$, $[U(1)_N]^3$, which are potentially troublesome. They are verified in Appendix \ref{appano}. Here, note that $\nu_{aR}$ as included from the outset are to cancel the gravity anomaly $[\mathrm{Gravity}]^2U(1)_N$ as well as the self-anomaly $[U(1)_N]^3$.          

A direct consequence of this note is that the often-studied 3-3-1 models are only self-consistent if they include $B-L$, thus $U(1)_N$, as a gauge symmetry. Otherwise, the 3-3-1 models are only effective theories at a low energy scale as often given in TeV range, for which $B-L$ acts as an approximate symmetry. And, the corresponding interactions that explicitly violate $B-L$ must present, in order to make the 3-3-1 models survival. All these will be proved in the next section, by verifying the unitarity argument of the current model and the 3-3-1 models.     

\subsection{Scalar sector, symmetry breaking, and $W$-parity} 

To break the 3-3-1-1 symmetry and generate the correct masses for the particles, we introduce the following scalars: 
\bea \eta &=&
\left(
\begin{array}{l}
\eta^{0,0}_1\\
\eta^{-1,0}_2\\
\eta^{q,n+1}_3
\end{array}\right)\sim \left(1,3,\fr{q-1}{3},\fr{n+1}{3}\right),\\
\rho &=&
\left(
\begin{array}{l}
\rho^{1,0}_1\\
\rho^{0,0}_2\\
\rho^{q+1,n+1}_3
\end{array}\right)\sim \left(1,3,\fr{q+2}{3},\fr{n+1}{3}\right),\\
\chi &=&
\left(
\begin{array}{l}
\chi^{-q,-n-1}_1\\
\chi^{-q-1,-n-1}_2\\
\chi^{0,0}_3
\end{array}\right)\sim\left(1,3,-\fr{2q+1}{3},-\fr 2 3 (n+1)\right),\\
\phi &\sim& (1,1,0,2),\eea where the superscripts denote $(Q,B-L)$ values respectively, while the subscripts indicate component fields under $SU(3)_L$. The scalars have such quantum numbers since $\eta$, $\rho$, $\chi$ couple a left-handed fermion to a corresponding right-handed fermion, whereas $\phi$ couples to $\nu_R\nu_R$ (as explicitly shown in the Yukawa Lagrangian below). Because $Q$ is conserved, only the electrically-neutral components $\eta_1$, $\rho_2$, $\chi_3$, $\phi$ can develop vacuum expectation values (VEVs), given by 
\bea \langle \eta \rangle &=& \fr{1}{\sqrt{2}}\left(
\begin{array}{c}
u \\
0\\
0
\end{array}\right),\hs
\langle \rho\rangle =
\fr{1}{\sqrt{2}} \left(
\begin{array}{c}
0\\
v \\
0
\end{array}\right),\hs
\langle \chi\rangle =
\fr{1}{\sqrt{2}} \left(
\begin{array}{c}
0\\
0\\
w
\end{array}\right),\hs \langle \phi\rangle = \fr{1}{\sqrt{2}} \La.\eea

The 3-3-1-1 symmetry is broken down to $SU(3)_C\otimes U(1)_Q\otimes U(1)_{B-L}$ due to $u,v,w$. Here, it undergoes two stages: the 3-3-1-1 symmetry to $SU(3)_C\otimes SU(2)_L\otimes U(1)_Y\otimes U(1)_{B-L}$ due to $w$, then $SU(3)_C\otimes SU(2)_L\otimes U(1)_Y\otimes U(1)_{B-L}$ to $SU(3)_C\otimes U(1)_Q\otimes U(1)_{B-L}$ due to $u,v$. Note that $u,\ v,\ w$ break only $N$, not $B-L$. Further, $\La$ breaks $B-L$, or $N$ totally since it also breaks $N$ in the previous stages, to a discrete symmetry, $U(1)_{B-L}\rightarrow P$ (shown below). In contrast to $Q$, the $B-L$ charge must be broken since its corresponding gauge boson should have a large mass, to escape from the detection. In summary, the gauge symmetry is broken as follows 
\be
SU(3)_C\otimes SU(3)_L\otimes U(1)_X\otimes U(1)_N \stackrel{u,v,w,\La}{\longrightarrow} SU(3)_C\otimes U(1)_Q\otimes P. 
\ee
The VEVs $w,\La$ provide the masses for the new particles, whereas $u,v$ are for those of the ordinary particles. To keep a consistency with the standard model, we assume $u,v\ll w,\La$.   
 
The charge $B-L=\beta' T_8 + N$ is the residual symmetry of $SU(3)_L\otimes U(1)_N$ since $[B-L]\langle \eta\rangle=[B-L]\langle \rho\rangle =[B-L]\langle \chi\rangle =0$ for $u,v,w\neq 0$. It transforms component fields/particles ($\Phi$) as \be \Phi\rightarrow \Phi'=U(\om)\Phi,\hs U(\om)=e^{i\om (B-L)},\ee where $\om$ is a transforming parameter. Further, $B-L$ is broken by $\langle \phi \rangle$ since $[B-L]\langle \phi\rangle = \sqrt{2}\La \neq 0$. Its remnant will conserve the vacuum, $U(\om) \langle \phi\rangle =\langle \phi \rangle$, i.e. $e^{i2\om} = 1$, and thus $\om = m \pi$ for $m=0,\pm1,\pm2,...$ We identify $P = e^{i\om (B-L)}= e^{im\pi (B-L)}= (-1)^{m(B-L)}$. Among such survival transformations, consider $m=3$, thus $P=(-1)^{3(B-L)}$, called matter parity. In addition, $P$ can be rewritten in a convenient form, \be P=(-1)^{3(B-L)+2s},\ee when multiplying the spin parity $(-1)^{2s}$, which is always conserved due to the angular momentum conservation. This is commonly known as $R$-parity, but in our case it results as a remnant of the gauge symmetry, given by \be P=(-1)^{3(\beta' T_8 + N)+2s}.\ee

If $k_a$ have ordinary $B-L$ numbers like those of the standard model, $n= \fr{2m-1}{3}=\pm \fr 1 3,\pm 1, \pm \fr 5 3,...$, it yields $P=1$ for all the fields of the model, which is trivial. The minimal versions belong to this case. However, since $k_a$ are new particles, we generally assume that $n$ is arbitrarily different from the ordinary ones, $n\neq \fr{2m-1}{3}$. Hence, the parity $P$ divides the model particles into two classes: 
\ben \item Normal particles: $P=1$. Include the standard model particles and some new ones: $\nu$, $e$, $u$, $d$, $\ga$, $W$, $Z$, $Z'$, $Z''$, $\eta_{1,2}$, $\rho_{1,2}$, $\chi_3$, $\phi$. They have ordinary $B-L$ numbers (or differ from these by even units as $\phi$ does), similarly to those of the standard model. They are even particles since $P=1$, as displayed in Table \ref{bang1}.        
\item Wrong particles: $P=P^+$ or $P^-$, where $P^\pm \equiv (-1)^{\pm(3n+1)}$. All the remaining particles, $k$, $j$, $X$, $Y$, $\eta_3$, $\rho_3$, $\chi_{1,2}$, have incorrect (wrong) $B-L$ numbers, in comparison to those of the standard model. They have a parity value of either $P^+$ or $P^-$, which is nontrivial due to $P^\pm \neq 1$, as shown in Table \ref{bang1}. Specially, the wrong particles become odd particles, i.e. $P=P^+=P^-=-1$, provided that $n=\fr{2m}{3}=0,\pm \fr 2 3,\pm \fr 4 3,\pm 2,...$   
 \een   
Therefore, the remarks are given in order \ben
\item $P$ is called $W$-parity, which distinguishes the wrong particles, called $W$-particles, from the normal (even) particles.  
\item Since $P$ is conserved, the $W$-particles are only coupled in pairs in interactions, which is analogous to superparticles in supersymmetry. Indeed, considering an interaction of $r+s$ $W$-fields, the $P$ conservation implies $(P^+)^r(P^-)^s=1$, where $r,s$ are integer, which happens if and only if $r=s$. The $P^+$ and $P^-$ fields always appear in pairs.   
\item Since $P$ is conserved, the lightest $W$-particle (LWP) is stabilized, which can be a dark matter candidate. The candidate must be electrically neutral, thus we have two dark matter models. (i) Model with $q=0$: the candidates are a fermion (as some combination of $k_a$), a gauge boson ($X_\mu$), and a scalar (as some combination of $\eta_3$ and $\chi_1$); (ii) Model with $q=-1$: the candidates are a gauge boson ($Y_\mu$) and a scalar (as some combination of $\rho_3$ and $\chi_2$).
\item Since $P$ is conserved, the $W$-scalars if being electrically neutral cannot develop VEVs. The VEVs as given above are unique. Also, there is no mixing between the $W$-particles and the normal particles, if they have the same electric charge. Here, the possible mixings are between exotic quarks and ordinary quarks as well as between new non-Hermitian gauge bosons and ordinary gauge bosons including $Z'$, $Z''$. Consequently, the dangerous tree-level flavor-changing neutral currents and $CP$ asymmetries due to such mixings are suppressed.            
\een
\begin{table}
\begin{tabular}{|c|ccccccccccccccccccccc|}
\hline
Particle & $\nu_a$ & $e_a$ & $u_a$ & $d_a$ & $\ga$ & $W$ & $Z$ & $Z'$ & $Z''$ & $\eta_{1,2}$ & $\rho_{1,2}$ & $\chi_3$ & $\phi$ & $k_a$ & $j_\al $ & $j_3$ & $X$ & $Y$ & $\eta_3$ & $\rho_3$ & $\chi_{1,2}$\\
\hline
$Q$ & 0 & $-1$ & $\fr 2 3$ & $-\fr 1 3$ & 0 & $1$ & 0 & 0 & 0 & $0,-1$ & $1,0$ & 0 & 0 & $q$ & $-\fr 1 3 -q$ & $\fr 2 3 + q$ & $-q$ & $-1-q$ & $q$ & $1+q$ & $-q,-1-q$\\
\hline  
$B-L$ & $-1$ & $-1$ & $\fr 1 3$ & $\fr 1 3$ & 0 & 0 & 0 & 0 & 0 & 0 & 0 & 0 & 2 & $n$ & $-\fr 2 3 -n$ & $\fr 4 3 +n$ & $-1-n$ & $-1-n$ & $1+n$ & $1+n$ & $-1-n$ 
\\
\hline
$P$ & 1 & 1 & 1 & 1 & 1 & 1 & 1 & 1 & 1 & 1 & 1 & 1 & 1 & $P^+$ & $P^-$ & $P^+$ & $P^-$ & $P^-$ & $P^+$ & $P^+$ & $P^-$\\
\hline
\end{tabular}
\caption{\label{bang1} The $Q$, $B-L$ charges and $W$-parity value for the model particles. The corresponding antiparticles have opposite $Q$ and $B-L$ charges, and $W$-parity conjugated.}
\end{table} 

\subsection{Total Lagrangian, fermion masses, and electric charge quantization} 

The total Lagrangian, up to the gauge fixing and ghost terms, is given by   
\bea \mathcal{L}&=&\sum_{\mathrm{fermion\ multiplets}} \bar{F}i\ga^\mu D_\mu F + \sum_{\mathrm{scalar\ multiplets}}(D^\mu S)^\dagger (D_\mu S) \crn
&&-\fr 1 4G_{i\mu\nu}G_i^{\mu\nu} -\fr 1 4  A_{i\mu\nu} A_i^{\mu\nu} -\fr 1 4 B_{\mu\nu}B^{\mu\nu} -\fr 1 4 C_{\mu\nu}C^{\mu\nu}\crn
&& + \mathcal{L}_{\mathrm{Yukawa}}- V(\eta,\rho,\chi,\phi),\eea  where the covariant derivative and field strength tensors are defined as   
\bea D_\mu &=& \pa_\mu + i g_s t_i G_{i\mu} + i g T_i A_{i\mu} + i g_X X B_\mu+ i g_N N C_\mu,\\
G_{i\mu\nu}&=&\pa_\mu G_{i\nu}-\pa_\nu G_{i\mu}-g_s f_{ijk} G_{j\mu} G_{k\nu},\\
A_{i\mu\nu}&=&\pa_\mu A_{i\nu}-\pa_\nu A_{i\mu}-g f_{ijk} A_{j\mu} A_{k\nu},\\
B_{\mu\nu}&=&\pa_\mu B_\nu-\pa_\nu B_\mu,\hs C_{\mu\nu}=\pa_\mu C_\nu-\pa_\nu C_\mu,\eea where $\{g_s,\ g,\ g_X,\ g_N\}$, $\{t_i,\ T_i,\ X,\ N\}$, and $\{G_i,\ A_i,\ B,\ C\}$ are coupling constants, generators, and gauge bosons of the 3-3-1-1 groups, respectively, and $f_{ijk}$ are $SU(3)$ structure constants.      

The Yukawa Lagrangian and scalar potential are obtained by
\bea \mathcal{L}_{\mathrm{Yukawa}}&=&h^\nu_{ab}\bar{\psi}_{aL}\eta\nu_{bR}+ h^e_{ab}\bar{\psi}_{aL}\rho e_{bR} + h^k_{ab}\bar{\psi}_{aL}\chi k_{bR}+h'^\nu_{ab}\bar{\nu}^c_{aR}\nu_{bR}\phi\crn
&& + h^j_{33}\bar{Q}_{3L}\chi j_{3R} + h^j_{\al \beta}\bar{Q}_{\al L} \chi^* j_{\beta R} + h^u_{3a} \bar{Q}_{3L}\eta u_{aR}\crn
&&+h^u_{\al a } \bar{Q}_{\al L}\rho^* u_{aR} +h^d_{3a}\bar{Q}_{3L}\rho d_{aR} + h^d_{\al a} \bar{Q}_{\al L}\eta^* d_{aR} +H.c.,\\ 
V(\eta,\rho,\chi,\phi) &=& \mu^2_1\eta^\dagger \eta + \mu^2_2 \rho^\dagger \rho + \mu^2_3 \chi^\dagger \chi + \mu^2_4 \phi^\dagger \phi +\la_1 (\eta^\dagger \eta)^2 + \la_2 (\rho^\dagger \rho)^2 \crn
&&+ \la_3 (\chi^\dagger \chi)^2 + \la_4 (\phi^\dagger \phi)^2 + \la_5 (\eta^\dagger \eta)(\rho^\dagger \rho) +\la_6 (\eta^\dagger \eta)(\chi^\dagger \chi)\crn
&&+\la_7 (\rho^\dagger \rho)(\chi^\dagger \chi)  +\la_{8} (\phi^\dagger \phi)(\eta^\dagger\eta)+\la_{9}(\phi^\dagger \phi)(\rho^\dagger \rho)+\la_{10}(\phi^\dagger \phi)(\chi^\dagger \chi)\crn
&& +\la_{11} (\eta^\dagger \rho)(\rho^\dagger \eta) +\la_{12} (\eta^\dagger \chi)(\chi^\dagger \eta)+\la_{13} (\rho^\dagger \chi)(\chi^\dagger \rho) + (\mu\eta \rho \chi +H.c.), 
\eea where the Yukawa couplings $h$'s and the scalar couplings $\la$'s are dimensionless, while $\mu_{1,2,3,4}$ and $\mu$ have the mass dimension. 

When the scalars develop VEVs, the fermions obtain masses. Conventionally, we write Dirac mass terms as $-\bar{f}_L m_f f_R+H.c.$ and Majorana mass terms as $-\fr 1 2 \bar{f}^c_{L,R} m^{L,R}_f f_{L,R}+H.c.$ The new fermions $k_a$ and $j_a$ possess $[m_k]_{ab}=-h^k_{ab}\fr{w}{\sqrt{2}}$, $[m_{j}]_{ab}=-h^j_{ab}\fr{w}{\sqrt{2}}$, with $h^j_{3\al}=h^j_{\al 3}=0$, which all have masses in $w$ scale. The masses of $e_a$, $u_a$ and $d_a$ are given by $[m_e]_{ab}=-h^e_{ab}\fr{v}{\sqrt{2}}$, $[m_u]_{3 a}=-h^u_{3 a}\fr{u}{\sqrt{2}}$, $[m_u]_{\al a}=h^u_{\al a}\fr{v}{\sqrt{2}}$, $[m_d]_{3 a}=-h^d_{3 a}\fr{v}{\sqrt{2}}$, and $[m_d]_{\al a}=-h^d_{\al a}\fr{u}{\sqrt{2}}$. Therefore, the ordinary charged-leptons and quarks gain the masses in the weak scales $u,v$, as usual. For the neutrinos, including the standard model $\nu_{aL}$ and their counterpart $\nu_{aR}$, we have Dirac masses $[m_\nu]_{ab}=-h^\nu_{ab}\fr{u}{\sqrt{2}}$ and Majorana masses $[m^R_\nu]_{ab}=-\sqrt{2}h'^\nu_{ab}\La$. Because of $u\ll \La$, the observed neutrinos $(\sim \nu_{aL})$ obtain masses via a type I seesaw mechanism, given by $m^L_\nu \simeq - m_\nu (m^R_\nu)^{-1} (m_\nu)^T\sim u^2/\La$, which is naturally small. Whereas, the heavy neutrinos $(\sim \nu_{aR})$ have masses $m^R_\nu$ as retained.   

\begin{figure}[h]
\begin{center}
\includegraphics{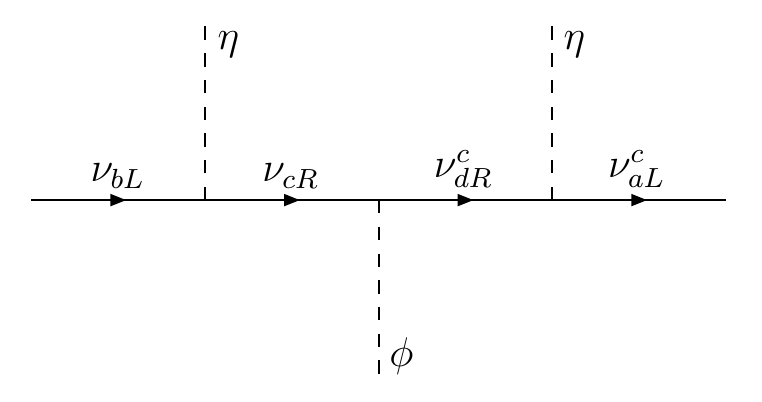}
\caption[]{\label{seesaw} Improved canonical-seesaw mechanism for neutrino masses.}  
\end{center}
\end{figure}

Indeed, such tiny masses for the neutrinos can be perturbatively (or dynamically) generated via a tree-level diagram in Fig. \ref{seesaw}, attached by three external Higgs fields $\eta$, $\phi$, $\eta$ with two respective internal lines $\nu_{R}$, $\nu^c_R$, when the electroweak and $B-L$ breakings happen, correspondingly determined by $\langle \eta\rangle $ and $\langle \phi\rangle $. The Majorana masses of $\nu_{aR}$ are also generated due to their interaction with $\phi$ (as the middle part in Fig. \ref{seesaw} graph) and the $B-L$ breaking by $\langle \phi \rangle$. The new observation is that the 3-3-1-1 symmetry suppresses all the neutrino mass types, but the electroweak and $B-L$ breakdown provides consistent masses for the neutrinos via such generalized Higgs mechanism. The type I seesaw mechanism is naturally recognized in this framework because it contains $\nu_{aR}$ as fundamental fermion constituents, and the Majorana masses are induced due to the $B-L$ gauge symmetry breaking. Further, this also works for a type II seesaw mechanism as mediated by a hypothetical scalar sextet (if one includes) that couples to $\psi_{aL}\psi_{bL}$ and to $\eta\eta\phi^*$. This contribution is just $\sim u^2/\La$ as the type I one is since the sextet mass is set by $\La$ scale. Both the mechanisms are corelated as achieved by the same symmetry breaking sources $\La$ and $u$.           

The standard model does not predict the electric charge quantization because of $Q=T_3+Y$, where the values of $T_3$ are quantized due to the non-Abelian nature of $SU(2)_L$ algebra, whereas the values of $Y$ are completely arbitrary. It is only chosen to describe the observed charges, does not explain them. The grand unified theories solve this issue since both $T_3$ and $Y$ are embedded in simple groups, thus the values of $Y$ are constrained due to the algebra structure. Our model provides an alternative solution, which is again due to the $B-L$ dynamics. For some pioneering works on the electric charge quantization, see \cite{addref1}.         

The electric charge operator is given by $Q=T_3+\beta T_8+X$, where $T_{3,8}$ are quantized due to the $SU(3)_L$ algebra structure. Therefore, $Q$ is quantized if $X$ for all multiplets is fixed. The ingredients in \cite{decq} are convenient for discussing further. First of all, the $X$-charges of $\eta$, $\rho$, $\chi$, and $\phi$ are constrained by $Q\langle \eta\rangle = Q\langle \rho \rangle =Q\langle \chi \rangle = Q\langle \phi \rangle=0$ because $Q$ is conserved. This gives $X_\phi=0$, while $X_{\eta,\rho,\chi}$ depend on $\beta$. The Yukawa Lagrangian is invariant under $U(1)_X$, which yields that all the right-handed fermions have $X$-charges related to those of the corresponding left-handed fermions and scalars. Also, the flavors $\psi_{aL}$ have the same $X$-charge, i.e. $X_{\psi_{1L}}=X_{\psi_{2L}}=X_{\psi_{3L}}\equiv X_{\psi_L}$, and this applies for other repetitive flavors such as $\nu_{aR}$, $e_{aR}$, $k_{aR}$, $Q_{\al L}$, $u_{aR}$, $d_{aR}$, and $j_{\al R}$, correspondingly. We denote $X_{f_{1R}}=X_{f_{2R}}=X_{f_{3R}}\equiv X_{f_{R}}$ ($f=\nu,\ e,\ k,\ u,\ d$), $X_{Q_{1L}}=X_{Q_{2L}}\equiv X_{Q_{\al L}}$, and $X_{j_{1 R}}=X_{j_{2 R}}\equiv X_{j_{\al R}}$. At this stage, we see that the charge of $Q_{3 L}$ is related to that of $Q_{\al L}$. Specially, we have the so-called quantization condition $X_{\nu_R}=-\fr 1 2 X_\phi=0$ due to the unique interaction of $\phi$ to $\nu_R \nu_R$. This leads to $X_{\psi_{L}}=X_\eta$ as fixed. The $[SU(3)_L]^2 U(1)_X$ anomaly cancelation gives $X_{Q_{\al L}}$ related to $X_{\psi_L}$ as fixed. Therefore, all $X$-charges are constrained, which most depend on $\beta$. Substituting into the electric charge operator, the ordinary particles have electric charges as observed, while the new particles have electric charges depending on $\beta$, i.e. $q$---the electric charge of $k_a$. Note that the electric charges of gauge bosons are always independent of $X$ and are either zero or fixed by $T_{3,8}$. If there is no $\nu_R\nu_R\phi$ interaction, the $X$-charges are unfixed, which are leaved as free parameters. Therefore, the $B-L$ dynamics is crucial to obtain the quantization of charges. The minimal versions have an different quantization condition \cite{decq}. 

Let us stress that the above ingredient (i.e. this model) explains only the electric charge quantization of ordinary particles. For the new particles as $k_a$, $j_\al$, $j_3$, $X$ and $Y$ bosons, and so on, their electric charges are not quantized, since $q$ (or $\beta$) is arbitrary.

\section{\label{gauge} Gauge bosons and constraints}

The mass Lagrangian of the gauge bosons is given by $\sum_{S=\eta,\rho,\chi,\phi}(D^\mu \langle S \rangle)^\dagger (D_\mu\langle S\rangle)$. We see that the gluons are always massless. The non-Hermitian gauge bosons $W$, $X$, and $Y$, which have been identified in Appendix \ref{appalge}, are physical particles with corresponding masses,
\be m^2_W=\fr{g^2}{4}(u^2+v^2),\hs m^2_X=\fr{g^2}{4}(w^2+u^2),\hs m^2_Y=\fr{g^2}{4}(w^2+v^2).\ee Here, $X$ and $Y$ are new gauge bosons, having large masses in $w$ scale, due to $w\gg u,v$. The $W$ field is identified as that of the standard model, which implies $u^2+v^2=(246\ \mathrm{GeV})^2$. The neutral gauge bosons $A_3$, $A_8$, $B$, and $C$ mix by themselves. However, it is easily to determine the photon, $Z$ boson, and new $\mathcal{Z}'$, given by   
\bea A &=& s_W A_3 + c_W \left(\beta t_W A_8 +\sqrt{1-\beta^2 t^2_W}B\right),\\
Z&=&c_W A_3 - s_W \left(\beta t_W A_8 +\sqrt{1-\beta^2 t^2_W}B\right),\\
\mathcal{Z}' &=&\sqrt{1-\beta^2 t^2_W} A_8 - \beta t_W B,\eea  where $s_W=e/g=t_X/\sqrt{1+(1+\beta^2)t_X^2}$, with $t_X=g_X/g$, is the sine of the Weinberg angle \cite{donglong}. Here, $\mathcal{Z}'$ is orthogonal to the field in the parentheses (i.e., to both $A$ and $Z$) that is coupled to the hypercharge $Y=\beta T_8 + X$, while $C$ is orthogonal to all $A$, $Z$, and $\mathcal{Z}'$. 

The photon $A$ is massless and decoupled (i.e. a physical particle) \cite{donglong}, while $Z$, $\mathcal{Z}'$ and $C$ mix. However, the mixing of $Z$ with the new $\mathcal{Z}'$ and $C$ is negligible due to the $\{u^2,v^2\}/\{w^2,\La^2$\} suppressions. Hence, the $Z$ boson can be considered as a physical particle with mass,
\be m^2_Z\simeq \fr{g^2}{4c^2_W}(u^2+v^2),\ee which is identical to that of the standard model. The fields $\mathcal{Z}'$ and $C$ finitely mix via a mass matrix as obtained by   
\bea \left(\begin{array}{cc}
m^2_{\mathcal{Z}'} & m^2_{\mathcal{Z}'C}\\
m^2_{\mathcal{Z}'C} & m^2_{C}\\
\end{array}\right),\eea where we have denoted $t_N=g_N/g$, and 
\be m^2_{\mathcal{Z}'} = \fr{g^2w^2}{3(1-\beta^2 t^2_W)},\hs m^2_{\mathcal{Z}'C}=-\fr{g^2 t_N\beta' w^2}{3\sqrt{1-\beta^2 t^2_W}},\hs m^2_{C}=4g^2 t_N^2 \La^2 + \fr 1 3 g^2t_N^2\beta'^2 w^2.\ee The $\mathcal{Z}'$-$C$ mixing angle is defined as 
\be t_{2\xi}=\fr{2m^2_{\mathcal{Z}'C}}{m^2_C-m^2_{\mathcal{Z}'}}=\fr{-2t_N\beta'\sqrt{1-\beta^2 t^2_W}w^2}{12t^2_N(1-\beta^2 t^2_W)\La^2+[t^2_N\beta'^2(1-\beta^2 t^2_W)-1]w^2}.\ee Therefore, the new neutral gauge bosons are 
\be Z'=c_\xi \mathcal{Z}'-s_\xi C,\hs Z''=s_\xi \mathcal{Z}'+c_\xi C,\ee with corresponding masses
\be m^2_{Z',Z''}=\fr 1 2 \left[m^2_{\mathcal{Z}'}+m^2_C\mp \sqrt{\left(m^2_{\mathcal{Z}'}-m^2_C\right)^2+4m^4_{\mathcal{Z}'C}}\right].\ee 

We rewrite $s^2_W=\fr{g^2_X}{g^2+(1+\beta^2)g^2_X}<\fr{1}{1+\beta^2}$. The model may encounter a Landau pole ($M$) at which $s^2_W(M)=\fr{1}{1+\beta^2}$ or $g_X(M)= \infty$. Hence, the model is consistent only if $M$ is larger than $w,\La$, and certainly it is larger than the weak scales $u,v$. We have a corresponding relation, $s^2_W(M)>s^2_W(u,v)$ since $g_X/g$ increases when the energy scale increases, which yields $|\beta|<\cot_W(u,v)\simeq 1.82455$ (for $s^2_W(u,v)\simeq 0.231$). With the aid of $\beta=-\fr{1+2q}{\sqrt{3}}$, we have $-2.08011 < q < 1.08011$. Therefore, the charge of $k_a$ is very constrained, and its bounds are very close to $-2$ and $1$, respectively. Demanding for integer $q$ yields only $q=-2,-1,0,1$. As a matter of fact, the model presents a low Landau pole of a few TeV for the bounds $q=1$ or $-2$ (see also \cite{landaupole}).                          

The quark flavors are nonuniversal under $SU(3)_L\otimes U(1)_X\otimes U(1)_N$ gauge symmetry because one generation of quarks transform differently from the two others, so there are FCNCs. Indeed, using $X=Q-T_3-\beta T_8$ and $N=B-L-\beta' T_8$, the interaction of neutral currents is given by 
\be \mathcal{L}_{\mathrm{NC}}=-g\bar{F}\ga^\mu[T_3 A_{3\mu} + T_8 A_{8\mu}+t_X(Q-T_3-\beta T_8)B_\mu+t_N (B-L-\beta' T_8)C_\mu]F,\ee where $F$ runs over all fermion multiplets of the model. It is easily realized that the leptons ($\nu_a,\ e_a,\ k_a$) and exotic quarks ($j_\al,\ j_3$) do not flavor-change. Also, the terms that contain $T_3$, $Q$, and $B-L$ do not lead to flavor-changing. The relevant part is 
\bea \mathcal{L}_{\mathrm{NC}}&\supset& -g\bar{q}_L\ga^\mu T_{8q}q_L (A_{8\mu}-\beta t_X B_\mu -\beta' t_N C_\mu)
=-\bar{q}_L\ga^\mu T_{8q} q_L\left(g'Z'_\mu + g''Z''_\mu\right), \eea
where $g'\equiv g(c_\xi/\sqrt{1-\beta^2 t^2_W}+s_\xi \beta' t_N)$, $g''\equiv g(s_\xi/\sqrt{1-\beta^2 t^2_W}-c_\xi \beta' t_N)$, and the field $q$ indicates to all ordinary quarks of either up type $q=(u_1,u_2,u_3)$ or down type $q=(d_1,d_2,d_3)$, with the corresponding $T_8$ values, $T_{8q}=\fr{1}{2\sqrt{3}}\mathrm{diag}(-1,-1,1)$. We change to mass basis, $q_{L,R}=V_{qL,qR}q'_{L,R}$, where $q'$ is either $q'=(u,c,t)$ or $q'=(d,s,b)$, and 
\bea \mathcal{L}_{\mathrm{NC}} &\supset& -\bar{q}'_L\ga^\mu (V^\dagger_{qL}T_{8q} V_{qL})q'_L(g'Z'_\mu+g''Z''_\mu),\crn
&\supset& -\fr{1}{\sqrt{3}}\bar{q}'_{iL}\ga^\mu q'_{jL} (V^*_{qL})_{3i}(V_{qL})_{3j}(g'Z'_\mu+g''Z''_\mu), \eea where the last one is FCNC Lagrangian, with $i\neq j$. This leads to the mixings of meson systems as described by the effective Lagrangian, 
\be \mathcal{L}^{\mathrm{eff}}_{\mathrm{FCNC}}=\fr 1 3 (\bar{q}'_{iL}\ga^\mu q'_{jL})^2 [(V^*_{qL})_{3i}(V_{qL})_{3j}]^2\left(\fr{g'^2}{m^2_{Z'}}+\fr{g''^2}{m^2_{Z''}}\right).\ee 

A strong bound comes from the $K^0-\bar{K}^0$ mixing, which constrains \cite{pdg}
\be \fr 1 3 [(V^*_{dL})_{31}(V_{dL})_{32}]^2\left(\fr{g'^2}{m^2_{Z'}}+\fr{g''^2}{m^2_{Z''}}\right)<\fr{1}{(10^4\ \mathrm{TeV})^2}. \ee Assuming that the up type quarks are flavor-diagonal, i.e. $V_{uL}=1$, the CKM matrix is just $V_{dL}$. We have $|(V^*_{dL})_{31}(V_{dL})_{32}|\simeq 3.6\times 10^{-4}$ \cite{pdg}, and thus 
\be \sqrt{\fr{g'^2}{m^2_{Z'}}+\fr{g''^2}{m^2_{Z''}}}<\fr{1}{2.078\ \mathrm{TeV}},\label{fcnc}\ee which directly implies $m_{Z'}>2.078\times g'\ \mathrm{TeV}$ and $m_{Z''}>2.078\times g''\ \mathrm{TeV}$. The new gauge bosons $Z'$, $Z''$ are in TeV range, provided that $g'$, $g''$ are in unity order.  

Another strong bound comes from the $B^0_s-\bar{B}^0_s$ mixing, given by \cite{pdg}  
\be \fr 1 3 [(V^*_{dL})_{32}(V_{dL})_{33}]^2\left(\fr{g'^2}{m^2_{Z'}}+\fr{g''^2}{m^2_{Z''}}\right)<\fr{1}{(100\ \mathrm{TeV})^2}.\ee
The CKM factor is $|(V^*_{dL})_{32}(V_{dL})_{33}|\simeq 3.9\times 10^{-2}$. Hence, we have 
\be \sqrt{\fr{g'^2}{m^2_{Z'}}+\fr{g''^2}{m^2_{Z''}}}<\fr{1}{2.25\ \mathrm{TeV}},\label{addfcncbb}\ee which leads to $m_{Z'}>2.25\times g'\ \mathrm{TeV}$ and $m_{Z''}>2.25\times g''\ \mathrm{TeV}$, slightly larger than the corresponding bounds obtained from the neutral kaon mixing.    

Further, without loss of generality, consider the first bound (\ref{fcnc}) for two cases [for the second bound (\ref{addfcncbb}), it can similarly be done]: 
\ben
\item $Z''$ is superheavy, i.e. $w\ll \La$. We have $m^2_{Z'}\simeq \fr{g^2w^2}{3(1-\beta^2 t^2_W)}$, $m^2_{Z''}\simeq 4g^2t^2_N \La^2$, and $\xi\simeq 0$. The condition (\ref{fcnc}) becomes \be \fr{1}{2.078\ \mathrm{TeV}}>\sqrt{\fr{3}{w^2}+\fr{\beta'^2}{4\La^2}}\simeq \fr{\sqrt{3}}{w} \longrightarrow w>3.6\ \mathrm{TeV}.\ee This is the common bound often derived for the 3-3-1 models, which is independent of $\beta$---the class of the 3-3-1 models. 
\item $Z''$ is comparable in mass to $Z'$, i.e. $w\sim \La$. The condition (\ref{fcnc}) leads to 
\be \left(\fr{1}{2.078\ \mathrm{TeV}}\right)^2> \fr{2|g'g''|}{m_{Z'}m_{Z''}}=\fr{6s_{2\xi}t_N\La\sqrt{3(1-\beta^2 t^2_W)}}{w^3}.\ee
For a simplicity, let us consider the $\mathcal{Z}'$-$C$ mixing to be maximal, i.e. $\xi=\pi/4$ or $\La/w=\sqrt{1-\beta'^2t^2_N(1-\beta^2 t^2_W)}/[2t_N\sqrt{3(1-\beta^2 t^2_W)}]$. The constraint becomes
\be w> 3.6\times [1-\beta'^2 t^2_N(1-\beta^2 t^2_W)]^{1/4}\ \mathrm{TeV}. \ee We have $w>3.57$ TeV and $\La=1.8w$ for the charges of $k_a$ as $(q,n)=(1,0),\ (-2,0)$, and $w>3.3$ TeV and $\La=0.5w$ for $(q,n)=(0,0)$. Here, we have taken $t_N=0.5$ and $s^2_W=0.231$.      
\een  

Finally, let us investigate the LEPII bounds for the process $e^+ e^-\rightarrow f\bar{f}$, where $f$ is an ordinary fermion, due to the exchange of new neutral gauge bosons such as $Z'$ and $Z''$ \cite{addcolliderbound}. The effective Lagrangian is therefore given by     
\be \mathcal{L}^{\mathrm{eff}}_{\mathrm{LEPII}}=\fr{g^2}{c^2_Wm^2_I}\left[\bar{e}\ga^\mu(a^I_L(e)P_L+a^I_R(e)P_R)e\right]\left[\bar{f}\ga_\mu(a^I_L(f)P_L+a^I_R(f)P_R)f\right],\ee where $I=Z',\ Z''$ and $a^I_{L,R}(f)=\fr 1 2 [g^I_V(f)\pm g^I_A(f)]$. Particularly considering $f=\mu$, we have 
\be \mathcal{L}^{\mathrm{eff}}_{\mathrm{LEPII}}= \fr{g^2}{c^2_W}\left(\fr{[a^{Z'}_L(e)]^2}{m^2_{Z'}}+\fr{[a^{Z''}_L(e)]^2}{m^2_{Z''}}\right)(\bar{e}\ga^\mu P_L e)(\bar{\mu}\ga_\mu P_L\mu)+(LR)+(RL)+(RR),\ee where the last three terms differ only from the first one in chiral structures, and 
\bea a^{Z'}_L (e) &=& \fr{c_W+\sqrt{3}\beta s_W t_W}{2\sqrt{3}\sqrt{1-\beta^2 t^2_W}}c_\xi +t_N c_W\left(1+\fr{\beta'}{2\sqrt{3}}\right)s_\xi,\crn a^{Z'}_R(e)&=&\fr{\beta s_W t_W}{\sqrt{1-\beta^2 t^2_W}}c_\xi + t_N c_W s_\xi,\\ 
a^{Z''}_{L,R}(e) &=& a^{Z'}_{L,R}(e)|_{c_\xi\rightarrow s_\xi,\ s_\xi\rightarrow -c_\xi}.\nn \eea 
Taking the typical bound as derived for $B-L$ gauge boson from \cite{addcolliderbound}, we get 
\be \fr{g^2}{c^2_W}\left(\fr{[a^{Z'}_L(e)]^2}{m^2_{Z'}}+\fr{[a^{Z''}_L(e)]^2}{m^2_{Z''}}\right) <\fr{1}{(6\ \mathrm{TeV})^2}.\ee Again, the masses of $Z',Z''$ are bounded in TeV range, $m_{Z'}>6\times \fr{g}{c_W}a^{Z'}_L(e)$ TeV and $m_{Z''}> 6 \times \fr{g}{c_W}a^{Z''}_L(e)$ TeV, assumed that $a^{Z',Z''}_L(e)$ are in unity order. To be concrete, consider that $Z''$ is superheavy, i.e. $w\ll \La$. With the aid of $\xi\simeq 0$ and $m^2_{Z'}\simeq g^2w^2/[3(1-\beta^2t^2_W)]$, we get 
\be w> 3\times (1+\sqrt{3}\beta t^2_W)\ \mathrm{TeV}.\ee Noting that $\beta=-(1+2q)/\sqrt{3}$, the bound for $w$ is 5.7 TeV, 3.9 TeV, 2.1 TeV, and 0.3 TeV, for $q=-2,-1,0,1$, respectively. The last case means that the $Z',\ Z''$ contributions are negligible since $w$ is in several TeV due to the other constraints aforementioned.                                 

\section{\label{unitaritys} Unitarity}

\begin{figure}[h]
\begin{center}
\includegraphics{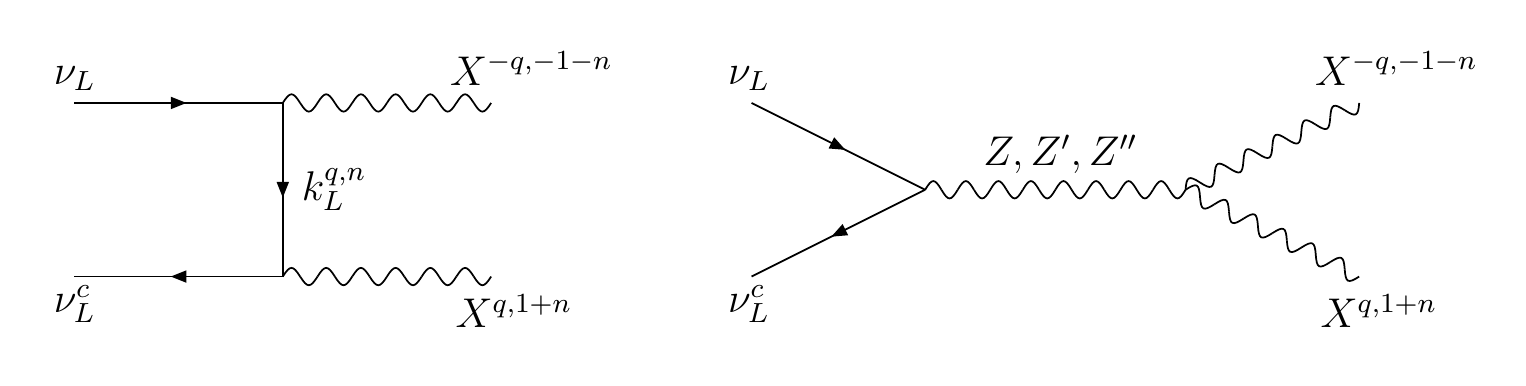}
\caption[]{\label{unitarity} Tree-level diagrams for $\nu^c_L\nu_L\rightarrow X^\dagger X$, where the $(Q,B-L)$ charges for $X$ and $k$ are explicitly displayed. We see that $Z''$ contributes since it interacts with $B-L$ charged particles such as $\nu_L$ and $X$. The remaining channels are identical to those in the 3-3-1 models.}  
\end{center}
\end{figure}

To investigate the unitarity of this model as well as of the previous proposals, it is enough to consider a high-energy scattering process of the standard model neutrinos ($\nu_L$) to the new gauge bosons ($X$): $\nu^c_L(p_1)\nu_L(p_2)\rightarrow X^\dagger(k_1) X(k_2)$. The tree-level contributions to the process are given in Figure \ref{unitarity}. The relevant interactions are obtained in Table \ref{dttd}.  
\begin{table}
\begin{tabular}{|c|c|}
\hline 
Vertex & Coupling \\
\hline 
$\bar{\nu}kX_\mu$ & $-\fr{ig}{\sqrt{2}}\ga^\mu P_L$ \\
$\bar{\nu}\nu Z_\mu$ & $-\fr{ig}{2c_W}\ga^\mu P_L$ \\
$\bar{\nu}\nu Z'_\mu$ & $-ig\left[\fr{1+\sqrt{3}\beta t^2_W}{2\sqrt{3}\sqrt{1-\beta^2 t^2_W}}c_\xi +\left(1+\fr{\beta'}{2\sqrt{3}}\right)t_N s_\xi\right]\ga^\mu P_L$\\
$\bar{\nu}\nu Z''_\mu$ & $-ig\left[\fr{1+\sqrt{3}\beta t^2_W}{2\sqrt{3}\sqrt{1-\beta^2 t^2_W}}s_\xi -\left(1+\fr{\beta'}{2\sqrt{3}}\right)t_N c_\xi\right]\ga^\mu P_L$ \\
$Z_\mu X^\dagger_\nu X_\al$ & $-\fr{ig}{2}\left(\sqrt{3}\beta s_W t_W -c_W\right)L^{\mu\nu\al}$\\
$Z'_\mu X^\dagger_\nu X_\al$ & $\fr{ig}{2}\sqrt{3(1-\beta^2 t^2_W)}c_\xi L^{\mu\nu\al}$ \\
$Z''_\mu X^\dagger_\nu X_\al$ & $\fr{ig}{2}\sqrt{3(1-\beta^2 t^2_W)}s_\xi L^{\mu\nu\al}$ \\
\hline 
\end{tabular}
\caption{\label{dttd} Relevant interactions for the $\nu^c_L \nu_L  \rightarrow X^\dagger X$ process. Here, $P_L=\fr 1 2 (1-\ga^5)$ and $L^{\mu\nu\al} = g^{\mu\nu}(q_1-q_2)^\al + g^{\nu \al}(q_2-q_3)^\mu+g^{\al \mu}(q_3-q_1)^\nu$, provided that all the momenta of $V_\mu(q_1)$, $X^\dagger_\nu(q_2)$, $X_\al (q_3)$ gauge bosons go into the vertex (otherwise, the signs of outgoing momenta are reversed).}
\end{table}

The amplitude of the $k_L$ exchange channel is computed as 
\bea iM(k_L) &=& \bar{v}(p_1)\left(-\fr{ig}{\sqrt{2}}\ga^\mu P_L\right)\fr{i}{\slashed{p}_2-\slashed{k}_2-m_k}\left(-\fr{ig}{\sqrt{2}}\ga^\nu P_L \right) u(p_2) \ep^*_\mu (k_1) \ep^*_\nu (k_2)\crn
&\simeq& \fr{ig^2}{2m^2_X}\bar{v}(p_1)\slashed{k}_1 P_L u(p_2).\eea Here, we have approximated $m_\nu, m_k\ll m_X$, and due to the high energy scattering, the longitudinal polarization components of the gauge bosons dominate, for which $\ep_\mu(k_{1})\simeq \fr{k_{1\mu}}{m_X}$ and $\ep_\nu(k_{2})\simeq \fr{k_{2\nu}}{m_X}$. The amplitude $M(k_L)$ is only one partial wave, but it is proportional to $s=4E^2$ at high energy. This violates the unitarity bound since the amplitude must be smaller than a constant.

The amplitudes of the $V=Z,Z',Z''$ exchange channels are obtained by 
\bea iM(V)=\bar{v}(p_1)(-if_V \ga^\al P_L) u(p_2) \fr{-i[g_{\al \la} - (k_1+k_2)_\al (k_1+k_2)_\la/m^2_V]}{(k_1+k_2)^2-m^2_V} (-i g_V L^{\la \mu\nu}) \ep^*_\mu (k_1) \ep^*_\nu (k_2),\nn\eea where $f_V$, $g_V$ stand for the coupling coefficients of $V$ to $\bar{\nu}\nu$ and $X^\dagger X$, respectively, which should be understood, and identified from Table \ref{dttd}. We approximate $L^{\la \mu\nu} \ep^*_\mu (k_1) \ep^*_\nu (k_2)\simeq (k_1-k_2)^\la k_1 k_2/m^2_X$ as $s$ is large. Hence, the amplitudes become 
\be iM(V) \simeq \fr{if_V g_V}{m^2_X}\bar{v}(p_1)\slashed{k}_1 P_L u(p_2). \ee             
For each $V$, it also goes as $s$, which violates the unitarity. 

Summing all the contributions, we have 
\be iM=iM(k_L)+iM(Z)+iM(Z')+iM(Z'')\sim \fr{1}{2}g^2+f_{Z} g_{Z}+f_{Z'} g_{Z'}+f_{Z''} g_{Z''}=0,\ee which exactly cancel out at high energy. The unitarity condition is satisfied by the 3-3-1-1 model. It is noteworthy that if the $Z''$ contribution is neglected, the unitarity is spoiled. Therefore, the 3-3-1 contributions by themselves violate the unitarity. The way for the 3-3-1 models to avoid this constraint is that $\xi=0$ or the $B-L$ breaking scale goes infinity, $\La=\infty$; otherwise it should include $B-L$ as a gauge symmetry. 

Note that a 3-3-1 model that regards $B-L$ as an approximate symmetry is only an effective theory because all $B-L$ violating interactions, which most include higher-dimensional ones, can enter as perturbations. Such theory loses predictive possibility as well as the unitarity is obviously violated at a high energy scale.

\section{\label{infbasym} Remarks on cosmological inflation and baryon asymmetry}

If the energy scale of $U(1)_N$ symmetry breaking happens at a very high scale like the grand unification one, the inflationary scenario which solves the difficulties of the hot big bang theory as well as quantum fluctuations in the inflating background can be obtained in this framework as linked (identical) to $U(1)_N$ breaking dynamics and driven by the $\phi$ potential. The $\phi$ field is inflaton. Because $U(1)_N$ is a local gauge symmetry, the radiative corrections to the inflaton potential include the interaction of inflation with $U(1)_N$ gauge boson ($Z''$) as well as the interactions of inflaton with the right-handed neutrinos ($\nu_{aR}$) and scalar triplets ($\eta,\rho,\chi$). Such interacting couplings are independent of details of $\beta$. Furthermore, due to the $W$-parity conservation, we can show that the physical scalar fields and their corresponding masses do not change when $\beta$ varies. That is to be said that the inflaton effective-potential and its consequences as given in the third article of \cite{d3311} generally apply for any $\beta$. Namely, the inflaton mass is in $10^{13}$ GeV order. The reheating temperature is either of $10^9$ GeV order if it dominantly decays into a pair of light Higgs bosons or a lower value if it dominantly decays into a pair of right-handed neutrinos. 

The baryon-number asymmetry of the universe can be obtained via lepogenesis processes due to the decays of $\nu_{aR}$. Again, the $U(1)_N$ breaking dynamics is crucial to generate the Majorana masses of $\nu_{aR}$ in order for a viable leptogenesis process. The $CP$ asymmetry decays of $\nu_{aR}$ proceed via two possibilities: (i) to ordinary charged leptons and corresponding charged scalars and (ii) to $k_a$ and a scalar combination of $\eta_3$ and $\chi_1$. The first case produces a baryon number as induced from the normal particle sector, whereas the second case generates a baryon number from the dark sector. If the right-handed neutrinos are produced as a result of the inflaton decays, we have the nonthermal leptogenesis processes. Otherwise, if the right-handed neutrinos are generated in the thermal bath of the universe (i.e. the inflaton dominantly decays into a pair of light Higgs bosons), we have the thermal leptogenesis sennario. Although, the general conclusions are similarly to the third article of \cite{d3311}, the details of the baryon number produced also depend on $\beta$ due to the contributions that come from interactions with the 3-3-1 gauge bosons.                                

\section{\label{conclusion} Conclusion} 

We have proved that the most economical gauge symmetry that supports the weak-isospin enlargement to $SU(3)_L$, with regarding electric charge and $B-L$ conservations, must be $SU(3)_C\otimes SU(3)_L\otimes U(1)_X\otimes U(1)_N$. The fact that the existence of $U(1)_N$ respecting $B-L$ symmetry is the same $U(1)_X$ for electromagnetic symmetry. The unification of electroweak and $B-L$ interactions is analogous to the Glashow-Weinberg-Salam criterion for weak and electromagnetic interactions. The theory predicts the fermion generation number and electric charge quantization, which all result from its consistent dynamics such as anomaly cancelation, QCD asymptotic freedom, and mass generation. Note that the electric charge quantization in the 3-3-1 models is only valid for those corresponding to the minimal versions as discussed. 

The most general fermion content has been introduced, which is independent of all the anomalies. The right-handed neutrinos $\nu_{aR}$ exist to cancel $B-L$ anomalies. The new fermions $k_a$, $j_a$ have general electric and $B-L$ charges and are related. The electric charge of $k_a$ is constrained by $-2.08011<q<1.08011$ (if integer charges assumed, it is $-2\leq q \leq 1$), while its $B-L$, $n$, is arbitrary, which has been assumed differently from ordinary ones [$n\neq (2m-1)/3$ for any integer $m$]. This suppresses the minimal versions. The third quark generation has been arranged differently from the first two under $SU(3)_L$. This leads to the bounds for  the new physics scales $w,\La$ in TeV range. Such different arrangement for the first or second generation instead is possible, but the new physics scales get a much higher bound, proportional to $10^3$ TeV (see, for a reference, \cite{ddrsp}).                               

The scalar sector has three triplets and one singlet, appropriately for the symmetry breaking and mass generation. A new $W$-parity (actually larger than $Z_2$) is recognized as a residual gauge symmetry. The wrong particles transform nontrivially under $W$-parity and are only coupled in pairs ($P^+,P^-$) in interactions. The normal particles which most are the standard model particles are $W$-parity even. There are two dark matter models corresponding to $q=0$ and $q=-1$. The previous analysis only realizes the former with $n=0$ \cite{d3311}. It is able to show that the neutral non-Hermitian gauge boson (either $X$ or $Y$) cannot be a dark matter since it completely annihilates before freeze-out. However, the fermion and scalar candidates are realistic since they can provide right abundance and relax search bounds. We have also shown that all the fermions and gauge bosons get consistent masses. The seesaw mechanisms responsible for small neutrino masses are naturally realized. The singlet scalar that breaks $B-L$ is crucial to determine the residual gauge symmetry, seesaw scales, charge quantization condition, and cosmological inflation \cite{d3311}.       

There is a finite mixing between the new neutral gauge bosons $\mathcal{Z}'$ and $C$. Unlike the new non-Hermitian gauge bosons $X,\ Y$, the physical new neutral gauge bosons $Z',\ Z''$ can interact with the ordinary fermions. The tree-level FCNCs due to $Z'$ and $Z''$ are bounded, yielding their masses in TeV range. The LEPII searches for $Z'$ and $Z''$ present similar bounds. Since the model does not induce proton decay despite the $B-L$ conservation, the $B-L$ breaking scale can be such low. There is no dangerous FCNC due to ordinary and exotic quark mixing since it is suppressed by $W$-parity. The unitarity of the model is verified. There are two folds for the 3-3-1 models, either they are this 3-3-1-1 model with $\La=\infty$ or they work as an effective theory at low energy respecting approximate $B-L$ symmetry.

\section*{Acknowledgments}

This research is funded by Vietnam National Foundation for Science and Technology Development (NAFOSTED) under grant number 103.01-2013.43.

\appendix 

\section{\label{appalge} Current algebra approach} 

The covariant derivative for $SU(3)_L$ is given by $D_\mu=\pa_\mu + ig T_i A_{i\mu}$, where $T_i$, $A_i$ and $g$ are the generators, gauge bosons and coupling constant, respectively. Let us work in the weak basis, which consists of the weight raising and lowering operators as well as the Cartan operators,        
\be T_\pm\equiv \fr{T_1\pm i T_2}{\sqrt{2}},\hs U_\pm\equiv \fr{T_4\pm i T_5}{\sqrt{2}},\hs V_\pm\equiv \fr{T_6\pm i T_7}{\sqrt{2}},\hs T_3,\hs T_8.\ee The corresponding gauge bosons are defined by 
\be W^{\pm,0}\equiv \fr{A_{1} \mp i A_{2}}{\sqrt{2}},\hs X^{\mp q,\mp(1+n)} \equiv \fr{A_{4} \mp i A_{5}}{\sqrt{2}},\hs Y^{\mp(1+ q),\mp(1 + n)} \equiv \fr{A_{6} \mp i A_{7}}{\sqrt{2}},\hs A_{3},\hs A_{8},\ee so that 
\be D_\mu = \pa_\mu + ig[(T_+ W^{+,0}_\mu + U_+ X^{-q,-1-n}_\mu + V_+ Y^{-1-q,-1-n}_\mu +H.c.)+ T_3A_{3\mu}+T_8A_{8\mu}]. \ee Above, the values superscripted to $W,\ X,\ Y$ denote $Q$ and $B-L$ charges respectively, whereas $A_{3,8}$ do not carry these charges. All these are obviously shown below.      

The gauge interactions of fermions arise from 
\bea \mathcal{L}= \bar{F}i\ga^\mu D_\mu F &\supset& (-g \bar{F}_L \ga^\mu T_+ F_L W^{+,0}_\mu -g \bar{F}_L \ga^\mu U_+ F_L X^{-q,-1-n}_\mu\crn
&& -g \bar{F}_L \ga^\mu V_+ F_L Y^{-1-q,-1-n}_\mu+H.c.)\crn
&&-g \bar{F}_L\ga^\mu T_3 F_L A_{3\mu}-g \bar{F}_L \ga^\mu T_8 F_L A_{8\mu},\eea where $F$ runs over all fermion multiplets of the model, and note that the generators vanish for $F_R$. Thus, the currents of $SU(3)_L$, which appear in the Lagrangian as $-gJ^\mu_A A_\mu$, can be read off, 
\be J^\mu_W=\bar{F}_L\ga^\mu T_+ F_L,\ J^\mu_X=\bar{F}_L\ga^\mu U_+ F_L,\ J^\mu_Y=\bar{F}_L\ga^\mu V_+ F_L,\ J^\mu_3=\bar{F}_L\ga^\mu T_3 F_L,\ J^\mu_8=\bar{F}_L\ga^\mu T_8 F_L. \ee This leads to the corresponding weak charges, 
\bea && T_+(t)\equiv \int d^3x J^0_W=\fr{1}{\sqrt{2}} \int d^3x (\nu^\dagger_{aL} e_{aL}+u^\dagger_{aL} d_{aL}),\crn
&&U_+(t)\equiv \int d^3x J^0_X=\fr{1}{\sqrt{2}}\int d^3x (\nu^\dagger_{aL} k_{aL}+u^\dagger_{3L} j_{3L}-j^\dagger_{\al L} d_{\al L}),\crn
&&V_+(t)\equiv\int d^3x J^0_Y=\fr{1}{\sqrt{2}}\int d^3 x (e^\dagger_{aL} k_{aL}+ d^\dagger_{3L} j_{3L}+ j^\dagger_{\al L} u_{\al L}),\crn
&& T_3(t)\equiv \int d^3x J^0_3=\fr 1 2 \int d^3x (\nu^\dagger_{aL} \nu_{aL}+u^\dagger_{aL} u_{aL}-e^\dagger_{aL} e_{aL}-d^\dagger_{aL} d_{aL}),\crn
&& T_8(t)\equiv \int d^3x J^0_8=\fr{1}{2\sqrt{3}}\int d^3 x (\nu^\dagger_{aL} \nu_{aL}+e^\dagger_{aL} e_{aL}-2k^\dagger_{aL} k_{aL} +u^\dagger_{3L} u_{3L}\crn
&&\hspace{3cm} +d^\dagger_{3L} d_{3L}-2j^\dagger_{3L} j_{3L}
-u^\dagger_{\al L} u_{\al L} -d^\dagger_{\al L} d_{\al L} +2j^\dagger_{\al L} j_{\al L}), 
\eea and $T_-(t)=[T_+(t)]^\dagger$, $U_-(t)=[U_+(t)]^\dagger$, $V_-(t)=[V_+(t)]^\dagger$. Here, the charges $(Q,\ B-L)$ for the new particles such as $X,\ Y,\ j_\al,\ j_3$ can be understood, provided that those charges for $k_a$ are $(q,n)$ as well as they are conserved.  

Using the canonical anticommutation relations for fermions, $\{f(\vec{x},t),f^\dagger(\vec{y},t)\}=\de^{(3)}(\vec{x}-\vec{y})$, we can check that the weak charges exactly satisfy the algebra of $SU(3)_L$ as usual. Particularly, the important commutation relations are 
\bea && [T_+(t),T_-(t)]=T_3(t),\crn 
 && [U_+(t),U_-(t)]=\fr 1 2 (T_3(t)+\sqrt{3}T_8(t)),\crn 
 && [V_+(t),V_-(t)]=\fr 1 2 (-T_3(t)+\sqrt{3}T_8(t)).\label{dssu3}\eea  

The $Q(t)$ and $[B-L](t)$ charges have the form,
\bea Q(t) &=& \int d^3x J^0_{em}=\int d^3x F^\dagger Q F\crn
&=&\int d^3x \left[-e^\+_{aL} e_{aL}+q k^\+_{aL}k_{aL}+(2/3) u^\+_{aL}u_{aL} -(1/3)d^\+_{aL}d_{aL}\right.\crn &&\left. +(2/3+q)j^\+_{3L} j_{3L}+(-1/3-q) j^\+_{\al L}j_{\al L}+(LL\rightarrow RR)\right],\eea\bea
[B-L](t) &=& \int d^3x J^0_{bl}=\int d^3x F^\dagger [B-L] F\crn
&=&\int d^3x \left[-\nu^\+_{aL}\nu_{aL}-e^\+_{aL} e_{aL}+n k^\+_{aL}k_{aL}+(1/3) u^\+_{aL}u_{aL} +(1/3)d^\+_{aL}d_{aL} \right.\crn &&\left. +(4/3+n)j^\+_{3L} j_{3L}+(-2/3-n) j^\+_{\al L}j_{\al L}+(LL\rightarrow RR)\right].
\eea
We see that $Q(t)$ and $[B-L](t)$ cannot, respectively, be any combination of $T_3(t)$ and $T_8(t)$ because $Q(t)$ and $[B-L](t)$ have the right currents as well (indeed, they are vectorlike). Therefore, the $SU(3)_L$ charges, $Q$ and $B-L$ do not form a closed algebra. Moreover, we derive \bea && [Q(t),T_{\pm}(t)]=\pm T_\pm,\hs [Q(t), U_{\pm}(t)]=\mp q U_\pm(t),\hs [Q(t), V_{\pm}(t)]=\mp (1+q) V_\pm(t),\\
&& [[B-L](t), U_{\pm}(t)]=\mp (1+n) U_\pm(t),\hs [[B-L](t), V_{\pm}(t)]=\mp (1+n) V_\pm(t),
\eea  which imply that the $SU(3)_L$ generators, $Q$ and $B-L$ do not commute. 

Putting $\beta=-(1+2q)/\sqrt{3}$, we deduce  
\bea Q(t)-T_3(t)-\beta T_8(t)&=&\int d^3x \left[\fr{-1+q}{3}\psi^\dagger_{aL}\psi_{aL}+\fr{1+q}{3}Q^\+_{3L}Q_{3L}-\fr{q}{3}Q^\+_{\al L}Q_{\al L}\right.\crn
&& -e^\dagger_{aR}e_{aR}+q k^\dagger_{aR} k_{aR}+\fr 2 3 u^\dagger_{aR}u_{aR}-\fr 1 3 d^\+_{aR}d_{aR}\crn
&&\left.+\left(\fr 2 3+q\right)j^\+_{3R}j_{3R}+\left(-\fr 1 3 -q\right)j^\+_{\al R}j_{\al R}\right] \equiv \int d^3x F^\dagger X F, 
\eea which defines a new Abelian charge, $X$, with the values for the multiplets, coinciding with those in (\ref{ddnng}). Surely, it is easily to check that the new charge $X(t)$ commutes with all the $SU(3)_L$ charges.  

Putting $\beta'=-2(1+n)/\sqrt{3}$, we obtain 
\bea [B-L](t)-\beta' T_8(t)&=&\int d^3x \left[\fr{-2+n}{3}\psi^\dagger_{aL}\psi_{aL}+\fr{2+n}{3}Q^\+_{3L}Q_{3L}-\fr{n}{3}Q^\+_{\al L}Q_{\al L}\right.\crn
&&-\nu^\dagger_{aR}\nu_{aR} -e^\dagger_{aR}e_{aR}+n k^\dagger_{aR} k_{aR}+\fr 1 3 u^\dagger_{aR}u_{aR}+\fr 1 3 d^\+_{aR}d_{aR}\crn
&&\left.+\left(\fr 4 3+n\right)j^\+_{3R}j_{3R}+\left(-\fr 2 3 -n\right)j^\+_{\al R}j_{\al R}\right] \equiv \int d^3x F^\dagger N F,\eea which yields another Abelian charge, $N$, with the values in agreement with (\ref{ddnng}). Also, $N(t)$ must commute with the $SU(3)_L$ charges.

Again, we conclude that the manifest gauge symmetry must be \be SU(3)_C\otimes SU(3)_L\otimes U(1)_X\otimes U(1)_N,\ee where $X$ and $N$, respectively, define the electric-charge and $B-L$ operators:
\be Q-T_3-\beta T_8 = X,\hs B-L-\beta' T_8=N.\ee

\section{\label{appano} Anomaly checking}

First note that $N=B-L-\beta' T_8$ and $X=Q-T_3-\beta T_8$. Furthermore, with the fermion content as given, the anomalies associated with $T_i$, $Q$ and $B-L$ always vanish. Therefore, the $X$ and $N$ anomalies are also cancelled. To see this explicitly, let us compute.    

\bea [SU(3)_C]^2U(1)_X&\sim& \sum_{\mathrm{quarks}} (X_{q_L}-X_{q_R}) = 3X_{Q_3}+2\times 3 X_{Q_\al}-3X_{u_a}-3X_{d_a}-X_{j_3}-2X_{j_\al}\crn
&=&3\left(1+q\right)/3+6\left(-q/3\right)-3\left(2/3\right)-3\left(-1/3\right)-\left(2/3+q\right)\crn
&&-2\left(-1/3-q\right)=0. \eea

\bea [SU(3)_C]^2U(1)_N&\sim& \sum_{\mathrm{quarks}} (N_{q_L}-N_{q_R}) = 3N_{Q_3}+2\times 3 N_{Q_\al}-3N_{u_a}-3N_{d_a}-N_{j_3}-2N_{j_\al}\crn
&=&3\left(2+n\right)/3+6\left(-n/3\right)-3\left(1/3\right)-3\left(1/3\right)-\left(4/3+n\right)\crn
&&-2\left(-2/3-n\right)=0. \eea

\bea 
[SU(3)_L]^2 U(1)_X &\sim& \sum_{\mathrm{(anti)triplets}} X_{F_L}= 3X_{\psi_a}+3X_{Q_3}+2\times 3 X_{Q_\al} \crn
&=& 3\left(-1+q\right)/3+3\left(1 +q\right)/3+6\left(-q/3\right)=0.  \eea 

\bea 
[SU(3)_L]^2 U(1)_N &\sim& \sum_{\mathrm{(anti)triplets}} N_{F_L}= 3N_{\psi_a}+3N_{Q_3}+2\times 3 N_{Q_\al} \crn
&=& 3\left(-2+n\right)/3+3\left(2 +n\right)/3+6\left(-n/3\right)=0.  \eea 

\bea [\mathrm{Gravity}]^2U(1)_X&\sim&\sum_{\mathrm{fermions}}(X_{f_L}-X_{f_R})=3\times 3 X_{\psi_a}+3\times 3 X_{Q_3}+2\times 3 \times 3 X_{Q_\al}\crn
&&-3\times 3 X_{u_a}-3\times 3 X_{d_a}-3X_{j_3}-2\times 3 X_{j_\al}-3X_{k_a}-3X_{e_a}-3X_{\nu_a}\crn
&=&3\times 3 (-1+q)/3+3\times 3 (1+q)/3+2\times 3\times 3 (-q/3)\crn
&&-3\times 3 (2/3)-3\times 3 (-1/3) -3(2/3+q)-2\times 3 (-1/3-q)\crn
&&-3q-3(-1)-3(0)=0.\eea

\bea [\mathrm{Gravity}]^2U(1)_N&\sim&\sum_{\mathrm{fermions}}(N_{f_L}-N_{f_R})=3\times 3 N_{\psi_a}+3\times 3 N_{Q_3}+2\times 3 \times 3 N_{Q_\al}\crn
&&-3\times 3 N_{u_a}-3\times 3 N_{d_a}-3N_{j_3}-2\times 3 N_{j_\al}-3N_{k_a}-3N_{e_a}-3N_{\nu_a}\crn
&=&3\times 3 (-2+n)/3+3\times 3 (2+n)/3+2\times 3\times 3 (-n/3)\crn
&&-3\times 3 (1/3)-3\times 3 (1/3) -3(4/3+n)-2\times 3 (-2/3-n)\crn
&&-3n-3(-1)-3(-1)=0.\eea

\bea [U(1)_X]^2U(1)_N&=&\sum_{\mathrm{fermions}}(X^2_{f_L}N_{f_L}-X^2_{f_R}N_{f_R})=3\times 3 X^2_{\psi_a}N_{\psi_a}+3\times 3 X^2_{Q_3} N_{Q_3}\crn &&+2\times 3\times 3 X^2_{Q_{\al}}N_{Q_{\al}}-3\times 3 X^2_{u_a}N_{u_a}-3\times 3 X^2_{d_a}N_{d_a}-3X^2_{j_3} N_{j_3}\crn
&&-2\times 3 X^2_{j_\al}N_{j_\al}-3X^2_{k_a} N_{k_a}-3X^2_{e_a} N_{e_a}-3X^2_{\nu_a}N_{\nu_a}\crn
&=&3\times 3 [(-1+q)/3]^2(-2+n)/3+3\times 3 [(1+q)/3]^2(2+n)/3\crn
&&+2\times 3\times 3 (-q/3)^2(-n/3)-3\times3(2/3)^2(1/3)-3\times 3(-1/3)^2(1/3)\crn
&&-3(2/3+q)^2(4/3+n)-2\times3(-1/3-q)^2(-2/3-n)-3q^2n\crn
&&-3(-1)^2(-1)-3(0)^2(-1)=0. \eea

\bea U(1)_X[U(1)_N]^2&=&\sum_{\mathrm{fermions}}(X_{f_L}N^2_{f_L}-X_{f_R}N^2_{f_R})=3\times 3 X_{\psi_a}N^2_{\psi_a}+3\times 3 X_{Q_3} N^2_{Q_3}\crn &&+2\times 3\times 3 X_{Q_{\al}}N^2_{Q_{\al}}-3\times 3 X_{u_a}N^2_{u_a}-3\times 3 X_{d_a}N^2_{d_a}-3X_{j_3} N^2_{j_3}\crn
&&-2\times 3 X_{j_\al}N^2_{j_\al}-3X_{k_a} N^2_{k_a}-3X_{e_a} N^2_{e_a}-3X_{\nu_a}N^2_{\nu_a}\crn
&=&3\times 3 [(-1+q)/3][(-2+n)/3]^2+3\times 3 [(1+q)/3][(2+n)/3]^2\crn
&&+2\times 3\times 3 (-q/3)(-n/3)^2-3\times3(2/3)(1/3)^2-3\times 3(-1/3)(1/3)^2\crn
&&-3(2/3+q)(4/3+n)^2-2\times3(-1/3-q)(-2/3-n)^2-3qn^2\crn
&&-3(-1)(-1)^2-3(0)(-1)^2=0. \eea 

\bea [U(1)_X]^3&=&\sum_{\mathrm{fermions}}(X^3_{f_L}-X^3_{f_R})=3\times 3 X^3_{\psi_a}+3\times 3 X^3_{Q_3}+2\times 3\times 3 X^3_{Q_\al}\crn
&&-3\times 3 X^3_{u_a}-3\times 3 X^3_{d_a}-3X^3_{j_3}-2\times 3 X^3_{j_\al}-3X^3_{k_a}-3X^3_{e_a}-3X^3_{\nu_a}\crn
&=&3\times 3 [(-1+q)/3]^3+3\times 3 [(1+q)/3]^3+2\times 3\times 3 (-q/3)^3\crn
&&-3\times3(2/3)^3-3\times 3(-1/3)^3-3(2/3+q)^3-2\times3(-1/3-q)^3\crn
&&-3q^3-3(-1)^3-3(-0)^3=0.\eea  

\bea [U(1)_N]^3&=&\sum_{\mathrm{fermions}}(N^3_{f_L}-N^3_{f_R})=3\times 3 N^3_{\psi_a}+3\times 3 N^3_{Q_3}+2\times 3\times 3 N^3_{Q_\al}\crn
&&-3\times 3 N^3_{u_a}-3\times 3 N^3_{d_a}-3N^3_{j_3}-2\times 3 N^3_{j_\al}-3N^3_{k_a}-3N^3_{e_a}-3N^3_{\nu_a}\crn
&=&3\times 3 [(-2+n)/3]^3+3\times 3 [(2+n)/3]^3+2\times 3\times 3 (-n/3)^3\crn
&&-3\times3(1/3)^3-3\times 3(1/3)^3-3(4/3+n)^3-2\times3(-2/3-n)^3\crn
&&-3n^3-3(-1)^3-3(-1)^3=0.\eea  

It is interesting that the anomalies are always cancelled, independent of $q$ and $n$, the corresponding $Q$ and $B-L$ charges of the new particles $k_a$. 

We see that although the $B,L$ symmetries are separately anomalous, taking $B-L$ into account makes the 3-3-1-1 model free from all the leptonic and baryonic anomalies \cite{d3311}.


\begin{thebibliography}{99}

\bibitem{pdg} K. A. Olive {\it et al.} (Particle Data Group), Chin. Phys. C {\bf 38}, 090001 (2014).

\bibitem{seesaw}
P. Minkowski, Phys. lett. B {\bf 67}, 421 (1977); M. Gell-Mann, P.
Ramond and R. Slansky, {\it Complex spinors and unified theories},
in {\it Supergravity}, edited by P. van Nieuwenhuizen and D. Z.
Freedman (North Holland, Amsterdam, 1979), p. 315; T. Yanagida, in
{\it Proceedings of the Workshop on the Unified Theory and the
Baryon Number in the Universe}, edited by O. Sawada and A.
Sugamoto (KEK, Tsukuba, Japan, 1979), p. 95; S. L. Glashow, {\it
The future of elementary particle physics}, in {\it Proceedings of
the 1979 Carg{\`e}se Summer Institute on Quarks and Leptons},
edited by M. L{\'e}vy et al. (Plenum Press, New York, 1980), pp.
687-713; R. N. Mohapatra and G. Senjanovi{\'c}, Phys. Rev. Lett.
{\bf 44}, 912 (1980); R. N. Mohapatra and G. Senjanovi{\'c}, Phys. Rev. D {\bf 23}, 165 (1981); G. Lazarides,
Q. Shafi and C. Wetterich, Nucl. Phys. B {\bf 181}, 287 (1981); J.
Schechter and J. W. Valle, Phys. Rev. D {\bf 25}, 774 (1982).

\bibitem{dmreview} See, for reviews, G. Jungman, M. Kamionkowski, and K. Griest, Phys. Rep. {\bf 267}, 195 (1996); G. Bertone, D. Hooper, and J. Silk, Phys. Rep. {\bf 405}, 279 (2005). 

\bibitem{d3311} 
P. V. Dong, T. D. Tham, and H. T. Hung, Phys. Rev. D {\bf 87}, 115003 (2013); P. V. Dong, D. T. Huong, F. S. Queiroz, and N. T. Thuy, Phys. Rev. D {\bf 90}, 075021 (2014); D. T. Huong, P. V. Dong, C. S. Kim, and N. T. Thuy, Phys. Rev. D {\bf 91}, 055023 (2015).

\bibitem{anomaly} D. J. Gross and R. Jackiw, Phys. Rev. D {\bf 6}, 477 (1972); H. Georgi and S. L. Glashow, Phys. Rev. D {\bf 6}, 429 (1972); J. Banks and H. Georgi, Phys. Rev. D {\bf 14}, 1159 (1976); S. Okubo, Phys. Rev. D {\bf 16}, 3528 (1977). 

\bibitem{nfg} P. H. Frampton, Phys. Rev. Lett. {\bf 69}, 2889 (1992). 

\bibitem{decq} P. V. Dong and H. N. Long, Int. J. Mod. Phys. A
{\bf 21}, 6677 (2006).

\bibitem{331m} F. Pisano and V. Pleitez, Phys. Rev.  D {\bf 46}, 410 (1992);
P. H. Frampton, in Ref. \cite{nfg}; R. Foot,
O. F. Hernandez, F. Pisano, and V. Pleitez, Phys. Rev. D {\bf 47},
4158 (1993).

\bibitem{331r} M. Singer, J. W. F. Valle, and J. Schechter, Phys.
Rev. D {\bf 22}, 738 (1980); J. C. Montero, F. Pisano, and V.
Pleitez, Phys. Rev. D {\bf 47}, 2918 (1993); R. Foot, H. N. Long,
and Tuan A. Tran, Phys. Rev. D {\bf 50}, R34 (1994). 

\bibitem{addref1} K. S. Babu and R. N. Mohapatra, Phys. Rev. Lett. {\bf 63}, 938 (1989); Phys. Rev. D {\bf 41}, 271 (1990); R. Foot, G. C. Joshi, H. Lew, and R. R. Volkas, Mod. Phys. Lett. A {\bf 5}, 2721 (1990); R. Foot, Mod. Phys. Lett. A {\bf 6}, 527 (1991); J. Sladkowski and M. Zralek, Phys. Rev. D {\bf 45}, 1701 (1992); M. Nowakowski and A. Pilaftsis, Phys. Rev. D {\bf 48}, 259 (1993); C. A. de S. Pires and O. P. Ravinez, Phys. Rev. D {\bf 58}, 035008 (1998).     

\bibitem{donglong} P. V. Dong and H. N. Long, Eur. Phys. J. C {\bf 42}, 325 (2005). 

\bibitem{landaupole} A. G. Dias, R. Martinez, and V. Pleitez, Eur. Phys. J. C {\bf 39}, 101 (2005).

\bibitem{addcolliderbound}  J. Alcaraz {\it et al.} (ALEPH, DELPHI, L3, OPAL Collaborations, LEP Electroweak Working Group), arXiv:hep-ex/0612034; M. Carena, A. Daleo, B. Dobrescu, and T. Tait, Phys. Rev. D {\bf 70}, 093009 (2004).

\bibitem{ddrsp} P. V. Dong, N. T. K. Ngan, and D. V. Soa, Phys. Rev. D {\bf 90}, 075019 (2014).

\end{thebibliography}
\end{document}